\documentclass[useAMS,usenatbib]{mn2e}
\usepackage{graphicx}

\title[Time-resolved observations of 5 \emph{INTEGRAL} CVs]{Time-resolved optical observations of five cataclysmic variables detected by \emph{INTEGRAL}}
\author[M.L. Pretorius]{Magaretha L. Pretorius\thanks{E-mail: retha@saao.ac.za} \\
South African Astronomical Observatory, PO Box 9, Observatory 7935, Cape Town, South Africa\\}
\begin{document}


\pagerange{\pageref{firstpage}--\pageref{lastpage}} \pubyear{}

\maketitle

\label{firstpage}

\begin{abstract}
The  ESA $\gamma$-ray telescope, \emph{INTEGRAL}, is detecting relatively more intrinsically rare cataclysmic variables (CVs) than were found by surveys at lower energies.  Specifically, a large fraction of the CVs that are \emph{INTEGRAL} sources consists of asynchronous polars and intermediate polars (IPs).  IP classifications have been proposed for the majority of CVs discovered by \emph{INTEGRAL}, but, in many cases, there is very little known about these systems.  In order to address this, I present time-resolved optical data of five CVs discovered through \emph{INTEGRAL} observations.  The white dwarf spin modulation is detected in high-speed photometry of three of the new CVs (IGR J15094-6649, IGR J16500-3307, and IGR J17195-4100), but two others (XSS J12270-4859 and IGR J16167-4957) show no evidence of magnetism, and should be considered unclassified systems.  Spectroscopic orbital period ($P_{orb}$) measurements are also given for IGR J15094-6649, IGR J16167-4957, IGR J16500-3307, and IGR J17195-4100.  
\end{abstract}


\section{Introduction}
Cataclysmic variable stars (CVs) are semi-detached interacting binaries, in which a white dwarf accretes from a low-mass, approximately main-sequence companion star.  CVs typically have orbital periods ($P_{orb}$) of hours.  Mass transfer is caused by orbital angular momentum loss, and usually leads to the formation of an accretion disc.  A comprehensive review of CVs may be found in \cite{bible}.

In about 20\% of CVs in the \cite{rkcat} catalogue, the white dwarf has a magnetic field sufficiently strong either to prevent the formation of a disc entirely, or at least disrupt it to a significant extent.  These are the magnetic CVs, and they are divided into two sub-types, namely polars and intermediate polars (IPs).  Polars show strong circular and linear polarization modulated at $P_{orb}$ (implying that the white dwarf rotation is synchronized with the binary orbit), while IPs are characterized by very stable pulsations, at periods $<P_{orb}$ in their X-ray and/or optical light curves.  Reviews of polars and IPs are given by \cite{Cropper90} and \cite{Patterson94}, respectively.  The distinction between the observational properties of polars and IPs is explained by a combination of white dwarf magnetic field strength and accretion rate, with polars having stronger magnetic fields and lower mass transfer rates ($\dot{M}$) than IPs.

Although technically the defining characteristic of a polar is white dwarf spin-orbit synchronism, several polars in fact rotate slightly asynchronously (see e.g.\ \citealt{StockmanSchmidtLamb88}; \citealt{CampbellSchwope99}).  The degree of asynchronism in these systems is small---the white dwarf spin period ($P_1$) and $P_{orb}$ differ by only $\sim1$\%---and they will attain spin-orbit synchronism on a short time-scale (e.g.\ \citealt{SchmidtStockman91}).

IPs are modelled as systems in which accretion is magnetically channelled from the inner edge of a truncated disc, onto a rapidly rotating white dwarf (see e.g.\ \citealt{BathEvansPringle74}; \citealt{PattersonRobinsonNather78}).  The defining short-period photometric pulsations are then identified with the rotation cycle of the white dwarf.  It is not uncommon for the dominant optical frequency to differ from the X-ray frequency; in those cases, the X-ray signal is believed to be the white dwarf spin frequency ($\omega$), and the optical frequency is most often $\omega - \Omega$, where $\Omega$ is the orbital frequency (i.e., the optical modulation arises from reprocessing of the anisotropic X-ray flux by a structure fixed in the reference frame rotating with the binary; e.g.\ \citealt{WarnerODonoghueFairall81}).  Other `orbital side bands', e.g.\ $\omega+\Omega$, can also be produced, and have been detected in some IPs (e.g.\ \citealt{Warner86}).  

Most IPs have harder X-ray spectra than polars, but both polars and IPs typically have large X-ray to optical flux ratios compared to non-magnetic CVs (e.g.\ \citealt{VerbuntBunkRitter97}).  High excitation lines, such as He\,{\scriptsize II}\,$\lambda$4686, are also usually prominent in the optical spectra of magnetic CVs.

The \emph{INTEGRAL} IBIS/ISGRI Soft Gamma-Ray Survey has for the first time systematically imaged a large fraction of the sky at energies above 20 keV (e.g.\ \citealt{Bird07}).  Roughly 5\% of sources detected in this survey are CVs, and a large fraction of these is made up of intrinsically rare systems.  Specifically, 2 \emph{INTEGRAL} CVs are asynchronous polars (there are only 6 known), and roughly half are IPs.  Overviews of the CV sample constructed from observations in this band may be found in \cite{BarlowKniggeBird06} and \cite{Revnivtsev08}.

Although \emph{INTEGRAL} is certainly detecting preferentially rare and interesting CVs, there is little information available on most of the systems that were unknown before their detection in this band.  Several of the CVs discovered by \emph{INTEGRAL} have been demonstrated to be IPs (e.g., \citealt{Araujo-Betancor03}; \citealt{Staude03}; \citealt{Gansicke05}; \citealt{Bonnet-Bidaud06}), but IP classifications are suggested for many more, on the basis of data that cannot cleanly distinguish IPs from other types of CVs.  In order to provide more reliable classifications, follow-up observations in the optical or softer X-ray bands are needed.  

I present time-resolved optical photometry and spectroscopy of five CVs discovered by \emph{INTEGRAL}.  I outline previously available information on the systems included in this study in Section~\ref{sec:targs}, and describe the new observations in Section~\ref{sec:obs} and \ref{sec:results}.  The results are discussed and summarized in Section~\ref{sec:disc} and \ref{sec:summ}.

\section{Target sample}
\label{sec:targs}

Five southern sources for which no time-resolved optical data exist were selected for this study; these CVs are listed in Table~\ref{tab:targs}.  XSS J12270-4859, IGR J15094-6649, IGR J16167-4957, and IGR J17195-4100 were identified as CVs by \cite{MasettiMorelliPalazzi06}, and IGR J16500-3307 by \cite{MasettiMasonMorelli08}.  Masetti et al.\ (2006, 2008) proposed that all 5 systems are IPs.

Broad spectral energy distributions (covering soft $\gamma$-ray to mid-IR wavelengths) of IGR J16167-4957 and IGR J17195-4100 are shown by \cite{TomsickChatyRodriguez06}, and \cite{Landi08} presents \emph{Swift} spectra of XSS J12270-4859, IGR J16167-4957, IGR J16500-3307, and IGR J17195-4100.

The \emph{RXTE} observations of \cite{Butters08} reveal candidates white dwarf spin modulations in XSS J12270-4859 and IGR J17195-4100.  This is discussed further in Sections~\ref{sec:12270} and ~\ref{sec:17195} below.

\begin{table*}
{\scriptsize
 \centering
  \caption{EW(H$\alpha$), $UBVRI$ magnitudes, and orbital and white dwarf spin periods for the targets in this study. 
}
  \vspace{0.2cm}
  \label{tab:targs}
  \begin{tabular}{@{}llllllllll@{}}
  \hline
Object          &\emph{ROSAT} counterpart&-EW(H$\alpha$)/\AA & $U$    & $B$    & $V$    & $R$    & $I$    & $P_{orb}/\mathrm{h}$ & $P_1/\mathrm{min}$ \\
  \hline
XSS J12270-4859 &1RXS J122758.8-485343   & 21(2)             & ---    & ---    & ---    & ---    &---     & ---                  & ---               \\
IGR J15094-6649 &1RXS J150925.7-664913   & 52.0(8)           &14.54(4)&15.28(2)&15.02(2)&14.72(1)&14.50(1)& 5.89(1)              & 13.4904(3)        \\
IGR J16167-4957 &1RXS J161637.2-495847   & 41(1)             &16.3(2) &17.13(8)&16.48(6)&16.01(4)&15.59(4)& 5.004(5)             & ---               \\
IGR J16500-3307 &1RXS J164955.1-330713   & 45.9(9)           &15.9(1) &16.49(5)&15.93(3)&15.51(2)&15.01(2)& 3.617(3)             &  9.9653(7)        \\
IGR J17195-4100 &1RXS J171935.6-410054   & 63.3(9)           &14.91(6)&15.64(2)&15.35(2)&14.98(2)&14.72(1)& 4.005(6)             & 18.9925(6)        \\
  \hline
  \end{tabular}
}
\end{table*}

\section{Observations}
\label{sec:obs}
Follow-up observations of the five CVs discussed in Section~\ref{sec:targs} were obtained from the Sutherland site of the South African Astronomical Observatory (SAAO).  

\subsection{$UBVRI$ photometry and low-resolution spectroscopy}
\label{sec:broadspec}
Broad, low-resolution spectra were obtained with the Grating Spectrograph on the SAAO 1.9-m telescope.  These observations were taken on the night starting 10 July 2008.  The no.\ 7 grating was used with a slit width of $1\farcs8$, yielding spectral resolution of $\simeq 5$~\AA\ over the wavelength range 3480 to 7280~\AA.  Object spectra were bracketed with arc lamp exposures to provide wavelength calibration, and flux calibration was achieved by observing spectrophotometric standard stars from \cite{StoneBaldwin83}, but, because of non-photometric conditions and slit losses, the absolute flux calibrations are not reliable. 

The CCD frames were processed and the spectra optimally extracted \citep{Horne86} using standard procedures in {\sc iraf}\footnote{{\sc iraf} is distributed by the National Optical Astronomy Observatories.}.  Wavelength calibration was performed by interpolating the dispersion solutions (from fits to the positions of the arc lines) of the arc lamp spectra taken immediately before and after each object spectrum.

\begin{figure}
 \includegraphics[width=84mm]{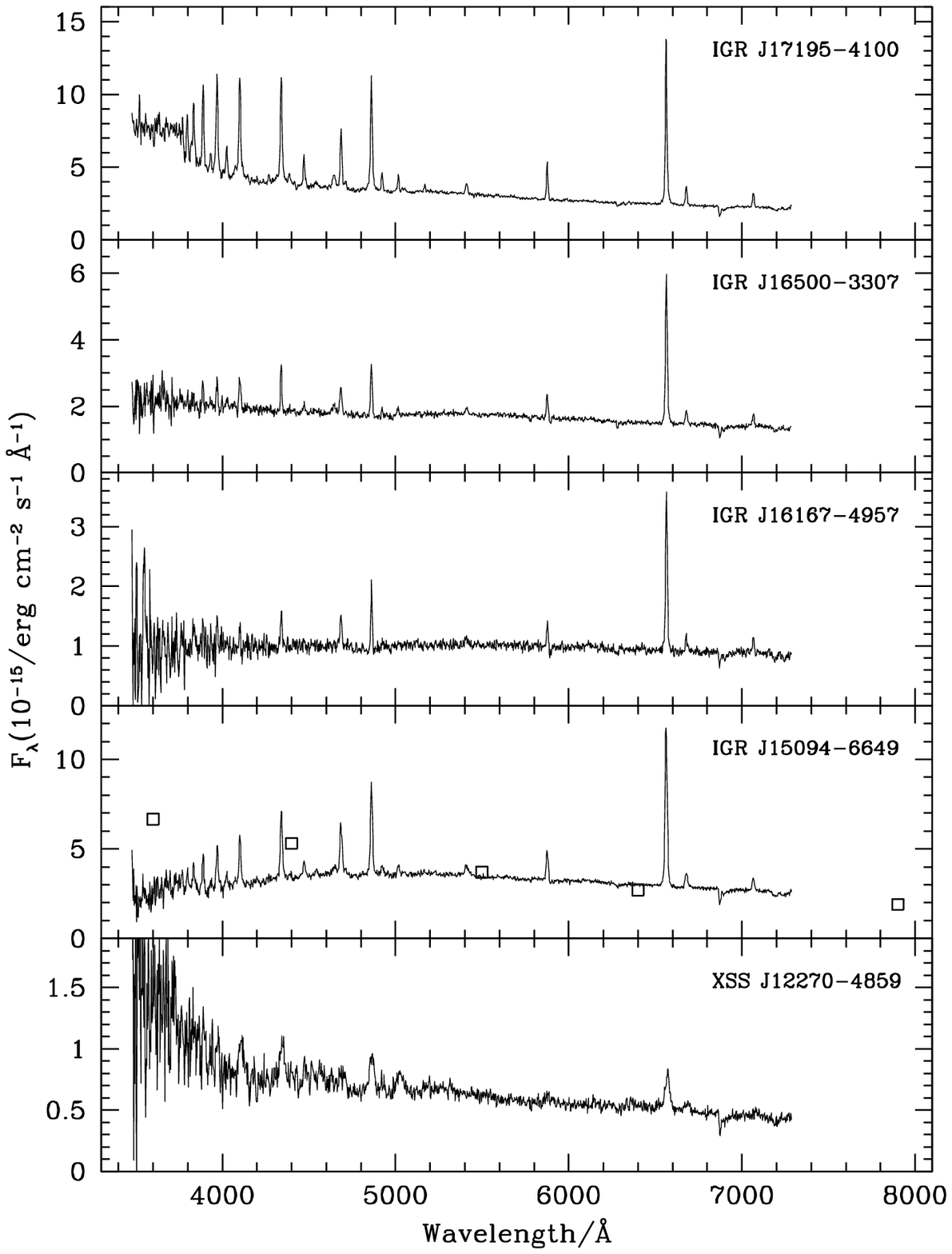}
 \caption{Low-resolution spectra of the CVs listed in Table~1, obtained with the SAAO 1.9-m telescope.  All spectra show Balmer, He\,{\scriptsize I}, and He\,{\scriptsize II} emission lines.  The C\,{\scriptsize III}/N\,{\scriptsize III}\,$\lambda\lambda$4640--4650 Bowen blend is also detected in the spectra of IGR J15094-6649, IGR J16500-3307, and IGR J17195-4100.  Flux measurements from the $UBVRI$ photometry, taken 30 days later, are superimposed on the spectrum of IGR J15094-6649; error bars are not shown, but are smaller than the symbols.
}
 \label{fig:lowresspec}
\end{figure}

The spectra are displayed in Fig.~\ref{fig:lowresspec}.  For IGR J15094-6649 and IGR J16167-4957, these are the averages of two 1000-s exposures, and for the remaining systems three individual spectra were averaged (individual exposure times were 900~s for XSS J12270-4859, and 1000~s for IGR J16500-3307 and IGR J17195-4100). The equivalent widths (EW) of the H$\alpha$ lines, listed in Table~\ref{tab:targs}, were measured from the averaged spectra; errors on these measurements were estimated using the method described by \cite{HowarthPhillips86}. 

$UBVRI$ photometry of four of the targets was obtained on the night starting 9 August 2008 (i.e.\ 30 days after the low-resolution spectra), using the University of Cape Town CCD photometer (the UCT CCD; see \citealt{uctccd}) on the SAAO 0.76-m telescope.  The photometry was performed using an adaptation of the program \textsc{dophot} \citep{dophot}.  $UBVRI$ magnitudes are listed in Table~\ref{tab:targs}.  Each of these measurements is the average of three exposures separated by a few minutes; photometric errors are given, but note that all these systems vary rapidly at larger amplitude (in white light), as a result of flickering and, in some cases, the white dwarf spin (see the light curves in Section~\ref{sec:results}).  

\subsection{High-speed photometry}
I took high-speed photometry of all five targets, using the SAAO 1-m and 0.76-m telescopes and the UCT CCD (this is a frame transfer CCD, implying that there is no dead time between exposures).  Table~\ref{tab:hs_phot} gives a log of the time-resolved photometry. 

These observations were unfiltered; the photometry therefore has a very broad bandpass and cannot be precisely placed on a standard photometric system.  The effective wavelength is similar to Johnson $V$, and the magnitude calibration approximates $V$ to within $\simeq 0.1$~mag.  

Most of the light curves are displayed in Section~\ref{sec:results}.  These are all differential light curves, implying that colour differences between the targets and the comparison stars were ignored in correcting the photometry for atmospheric extinction.  

\begin{table*}
 \centering
 \caption{Log of the high-speed photometry.  Dates are for the start of the night, and HJD for the middle of the first exposure.  The average magnitude for each run is listed in the final column. $t_{int}$ is the integration time (and also the time resolution), and `:' denotes an uncertain value.}
  \label{tab:hs_phot}
  \begin{tabular}{@{}llllllll@{}}
  \hline
Object    & Run no. & Date           & HJD $2454000.0 +$ & Length/h & $t_{int}$/s & telescope & $V$ \\
  \hline
XSS J12270-4859 & RP55 & 2008 Apr 23 & 580.2285553 & 4.6 & 10 & 1-m    & 16.1 \\
                & RP57 & 2008 Apr 24 & 581.2155297 & 2.4 &  8 & 1-m    & 16.1 \\
                & RP69 & 2008 Apr 29 & 586.2209495 & 4.0 &  8 & 1-m    & 16.1 \\[0.1cm]
IGR J15094-6649 & RP62 & 2008 Apr 26 & 583.3407779 & 1.6 &  8 & 1-m    & 14.7 \\
                & RP64 & 2008 Apr 27 & 584.3210149 & 3.9 &  7 & 1-m    & 14.6 \\
                & RP67 & 2008 Apr 28 & 585.2980160 & 3.1 &  7 & 1-m    & 14.5:\\
                & RP70 & 2008 Apr 29 & 586.4190670 & 6.3 &  7 & 1-m    & 14.7 \\[0.1cm]
IGR J16167-4957 & RP56 & 2008 Apr 23 & 580.4320679 & 5.8 & 10 & 1-m    & 16.5 \\
                & RP59 & 2008 Apr 24 & 581.3929664 & 3.4 & 10 & 1-m    & 16.5 \\
                & RP72 & 2008 Aug  2 & 681.2551003 & 4.3 & 10 & 0.76-m & 16.5:\\[0.1cm]
IGR J16500-3307 & RP65 & 2008 Apr 27 & 584.4930975 & 2.2 &  8 & 1-m    & 15.9 \\
                & RP68 & 2008 Apr 28 & 585.4319068 & 1.6 &  8 & 1-m    & 15.9:\\
                & RP71 & 2008 May  4 & 591.4937969 & 3.1 & 10 & 0.76-m & 15.9 \\
                & RP73 & 2008 Aug  3 & 682.2059835 & 1.7 & 10 & 0.76-m & 15.9:\\
                & RP74 & 2008 Aug  4 & 683.2085064 & 6.1 & 10 & 0.76-m & 16.0 \\
                & RP76 & 2008 Aug  5 & 684.2065262 & 3.7 & 10 & 0.76-m & 16.0 \\[0.1cm]
IGR J17195-4100 & RP77 & 2008 Aug  5 & 684.3657404 & 3.3 &  8 & 0.76-m & 15.2 \\
                & RP79 & 2008 Aug  6 & 685.2031547 & 7.2 &  8 & 0.76-m & 15.2 \\
                & RP80 & 2008 Aug  7 & 686.2050704 & 7.1 &  8 & 0.76-m & 15.2 \\
                & RP82 & 2008 Aug  8 & 687.2075755 & 6.8 &  8 & 0.76-m & 15.1 \\
                & RP84 & 2008 Aug  9 & 688.3490611 & 3.2 &  8 & 0.76-m & 15.1 \\
                & RP86 & 2008 Aug 10 & 689.2999440 & 4.7 &  8 & 0.76-m & 15.2 \\[0.1cm]
  \hline
  \end{tabular}
\end{table*}

\subsection{Time-resolved spectroscopy}
In order to measure orbital periods, I obtained phase-resolved medium-resolution spectra of 4 systems (IGR J15094-6649, IGR J16167-4957, IGR J16500-3307, and IGR J17195-4100) with the SAAO 1.9-m telescope and the Grating Spectrograph.  Grating no.\ 6 and a slit width of $1\farcs5$ were used, resulting in $\simeq 2\,\mathrm{\AA}$ resolution over the wavelength range 5220 to 6960~\AA.  Details of the time-resolved spectroscopy are given in Table~\ref{tab:tr_spect}.  

Regular arc lamp exposures were taken to provide wavelength calibration.  The data were reduced in the same way as described in Section~\ref{sec:broadspec} above.

\begin{table}
{\footnotesize
 \centering
  \caption{Log of the time-resolved spectroscopy.  Dates are for the start of the night; the year is 2008 in all cases. HJD is for the middle of the first exposure. $t_{int}$ denotes the integration time. }
  \label{tab:tr_spect}
  \begin{tabular}{@{}lllll@{}}
  \hline
Object          & Date        & HJD $2454000.0 +$ & $t_{int}$/s & no.\ of \\
                &             &                   &             & spectra\\
  \hline
IGR J15094-6649 & Jul 11 & 659.21746424 &  800,600 & 28 \\
                & Jul 12 & 660.19628522 &  900,800 & 39 \\
                & Jul 13 & 661.19655116 &  600     & 21 \\[0.1cm]
IGR J16167-4957 & Jul 28 & 676.20795926 & 1000     & 22 \\
                & Jul 29 & 677.20722879 & 1000     & 14 \\
                & Jul 30 & 678.21737633 & 1000     & 19 \\
                & Aug  1 & 680.21270551 & 1000     & 22 \\[0.1cm]
IGR J16500-3307 & Jul 20 & 668.21258923 & 1000,900 & 23 \\
                & Jul 21 & 669.21261534 &  900     & 27 \\
                & Jul 22 & 670.21869310 &  900     & 22 \\
                & Jul 23 & 671.21221271 &  900     & 23 \\[0.1cm]
IGR J17195-4100 & Jul 24 & 672.21196843 &  600     & 31 \\
                & Jul 25 & 673.21341216 &  600     & 44 \\
                & Jul 26 & 674.21225069 &  600     & 29 \\
  \hline
  \end{tabular}
}
\end{table}

Radial velocities were computed with the Fourier cross correlation method described by \cite{fxcor}, as implemented in the {\sc fxcor} routine in {\sc iraf}.  The radial velocities were found by correlating spectra with a template (made, over a few iterations, by shifting all individual spectra of a given system to 0 velocity, using the radial velocity measurement from the previous iteration, and averaging them).  The wavelength range of the spectra covers H$\alpha$ and the He\,{\scriptsize I}\,$\lambda$5876 and $\lambda$6678 lines, but only the H$\alpha$ lines were included in the correlation.  In the case of IGR J15094-6649, the cross correlation with a template spectrum did not give satisfactory results, probably because the H$\alpha$ emission line profile changes with orbital phase.  For this system, I therefore measured radial velocities using the H$\alpha$ line wings with the double Gaussian technique of \cite{SchneiderYoung80}, as implemented in the program {\sc molly}, written by Tom Marsh.  

\begin{figure}
 \includegraphics[width=84mm]{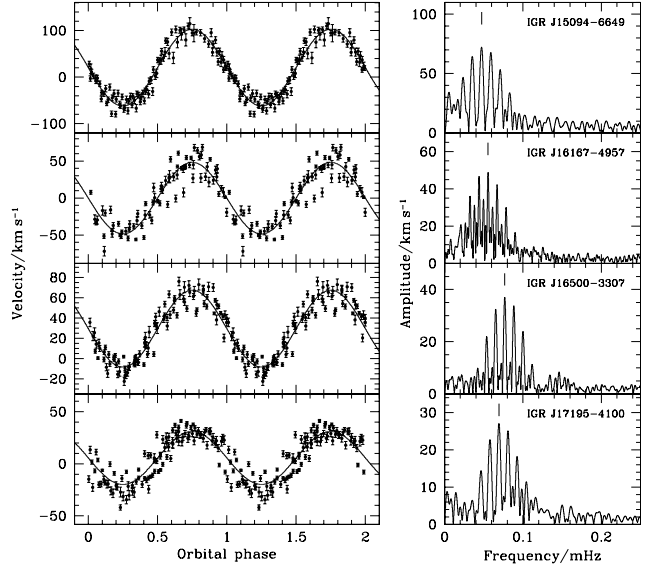}
 \caption {Fourier transforms and phase-folded radial velocity curves, together with least squares sine fits.  Vertical bars in the Fourier transforms mark the periods used to fold the data.  One cycle is repeated in the folded radial velocity curves.  
}
 \label{fig:rvandfts}
\end{figure}

A function of the form 
\begin{equation}
v(t)=\gamma - K \sin \left [ 2\pi \left ( t-T_0\right )/P_{orb} \right ]
\label{eq:rvfit}
\end{equation}
was fitted to the radial velocity curves by least squares ($T_0$ is the epoch of red to blue crossing of radial velocity). The phase-folded radial velocities with best-fit sinusoids over plotted, as well as Fourier transforms of the radial velocity curves, are shown in Fig.~\ref{fig:rvandfts}. Best-fit parameters are given in Table~\ref{tab:rvfit}.
The formal errors in radial velocities appear to be too small (see the left-hand panels of Fig.~\ref{fig:rvandfts}), but this is not a serious concern, since the amplitudes of the orbital modulations are large compared to the uncertainty indicated by the scatter in radial velocity around these modulations.
The frequency identified with the orbital modulation is marked by a vertical bar in each Fourier transform, and was used to phase-fold the data.  It was in all cases possible to determine the cycle count between radial velocity curves from different nights unambiguously\footnote{The errors in radial velocity were adjusted by assuming intrinsic scatter that results in reduced $\chi^2=1$ for the best fit to the data.  Using the same total errors, I then find that fits with periods near other aliases in the Fourier transforms all produce values of $\chi^2$ for which the probability of exceeding it is $< 0.001$.  Therefore, all except the strongest alias in each Fourier transform can be rejected.}.

\begin{table}
{\footnotesize
 \centering
  \caption{Parameters of the radial velocity fits (see equation~\ref{eq:rvfit}).}
  \label{tab:rvfit}
  \begin{tabular}{@{}lllll@{}}
  \hline
Object          & $\gamma/\mathrm{km\,s^{-1}}$& $K/\mathrm{km\,s^{-1}}$& $T_0-2454000.0$& $P_{orb}$/d \\
  \hline
IGR J15094-6649 & 20(2)                      & 82(2)                  & 659.243(1)     & 0.2453(4) \\
IGR J16167-4957 & 0(2)                       & 48(2)                  & 676.317(2)     & 0.2085(2) \\
IGR J16500-3307 & 29.5(9)                    & 37(1)                  & 668.2192(8)    & 0.1507(1) \\
IGR J17195-4100 & 4.4(9)                     & 28(1)                  & 672.306(1)     & 0.1669(3) \\
  \hline
  \end{tabular}
}
\end{table}

\begin{figure*} 
 $\begin{array}{c@{\hspace{5mm}}c@{\hspace{5mm}}c@{\hspace{5mm}}c}
 \includegraphics[width=36mm]{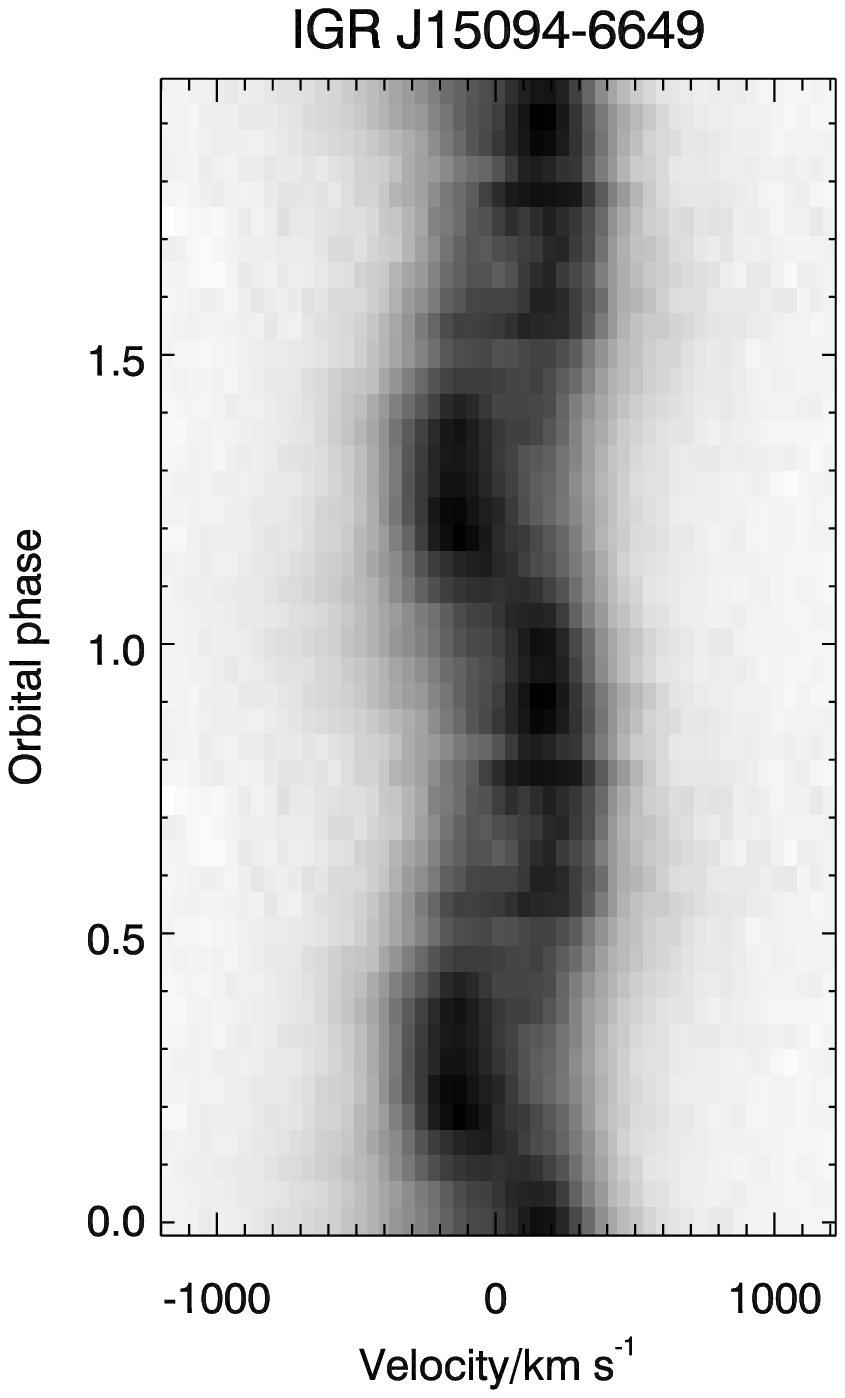} &
 \includegraphics[width=36mm]{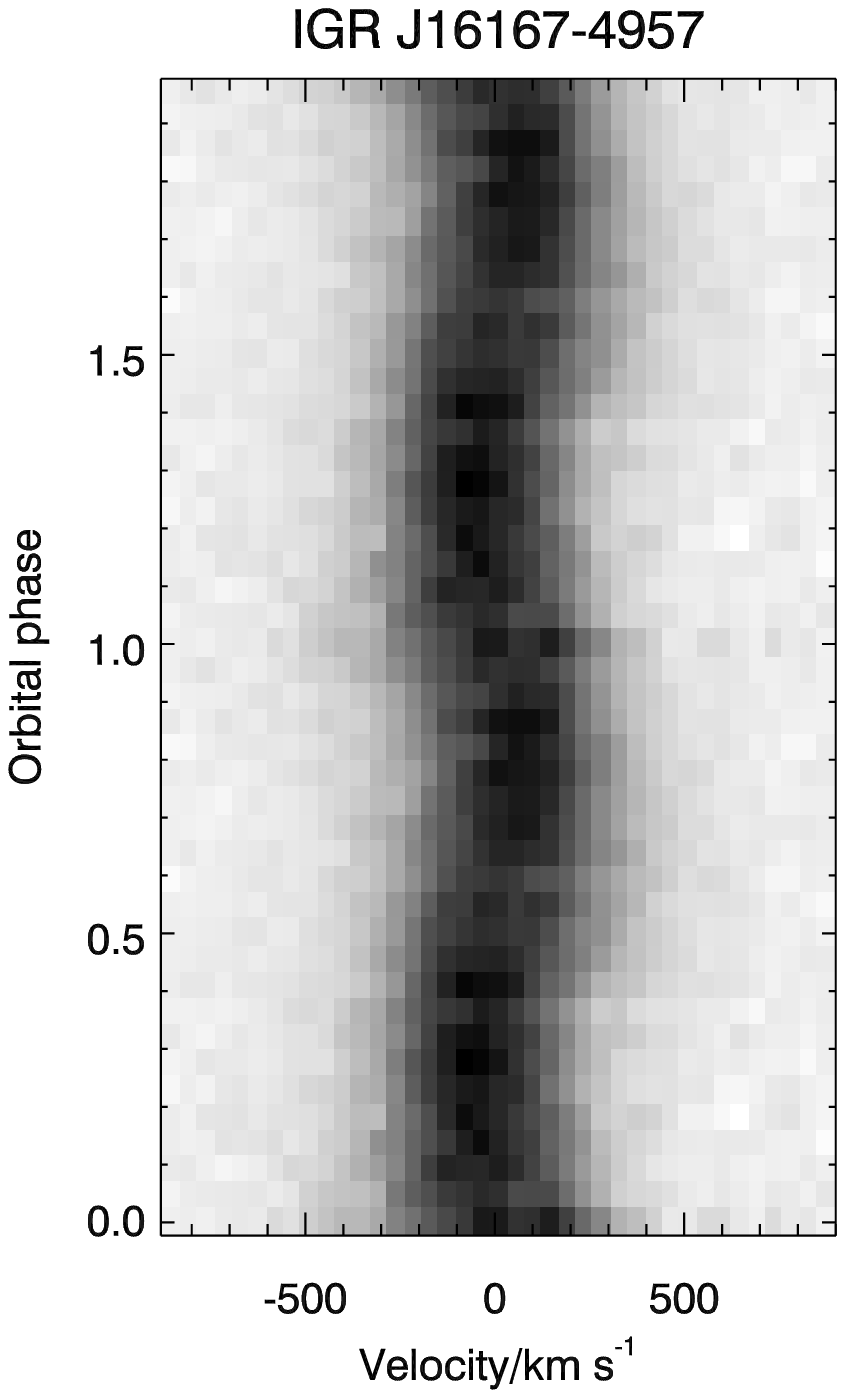} &
 \includegraphics[width=36mm]{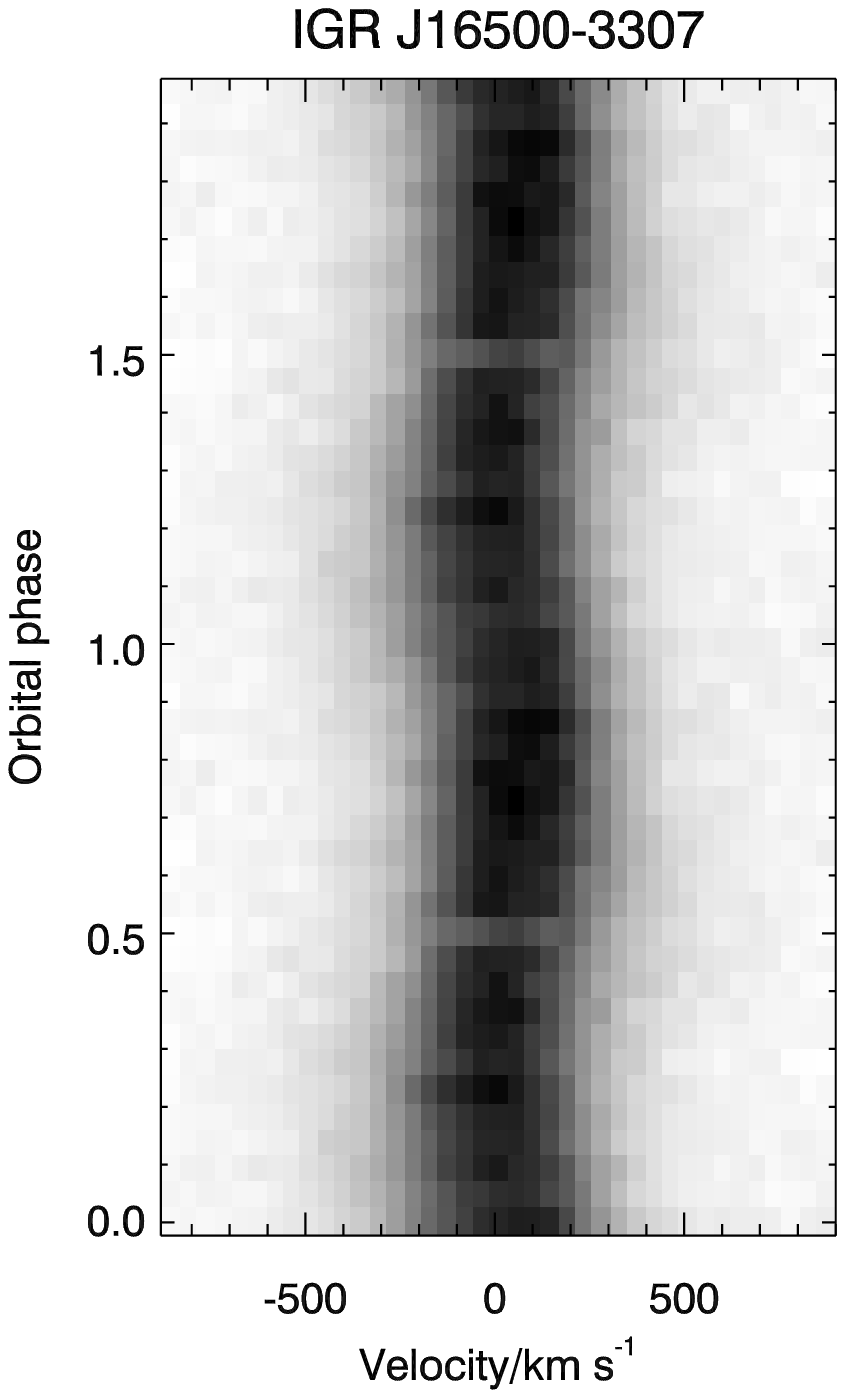} &
 \includegraphics[width=36mm]{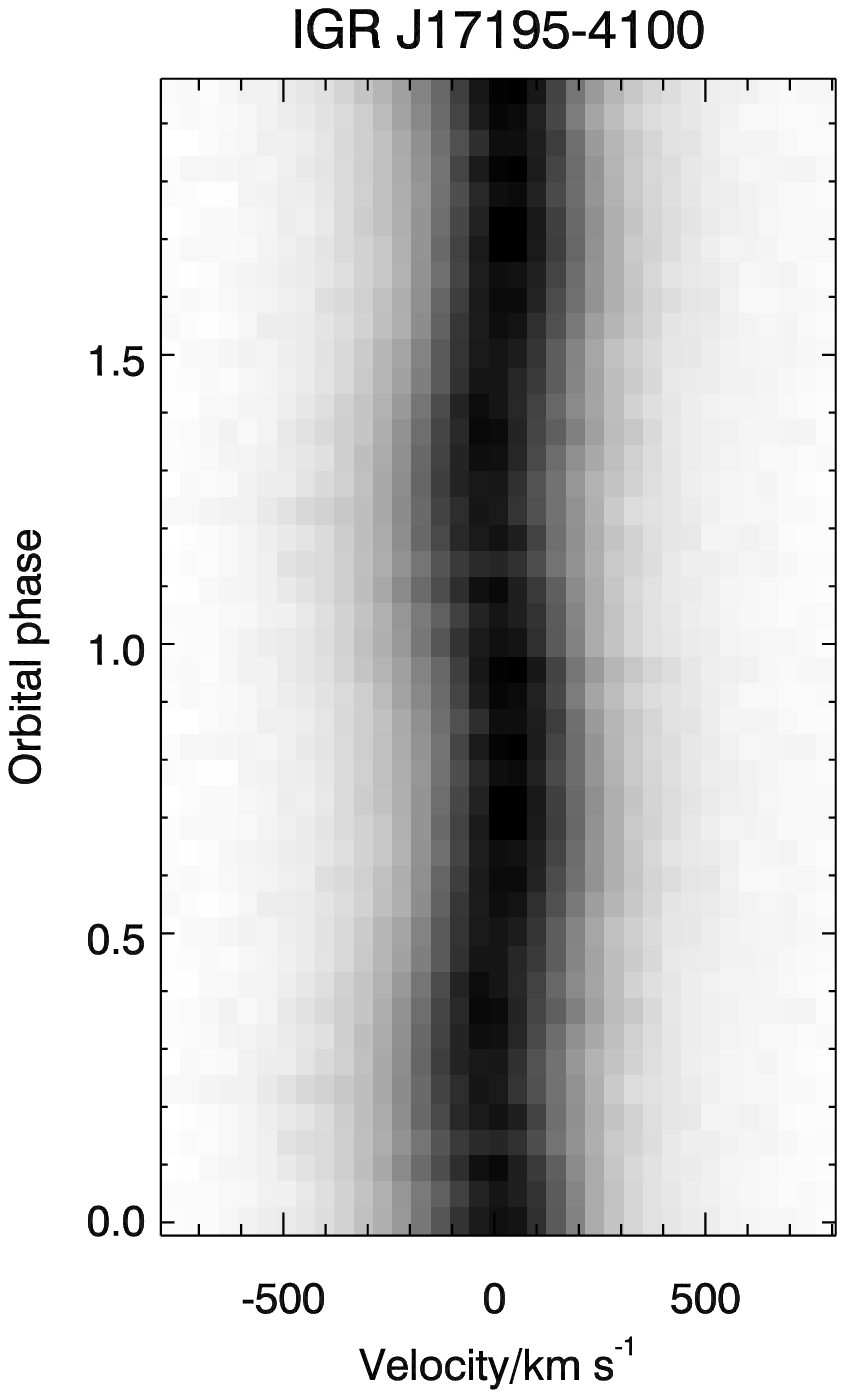}  
 \end{array}$
 \caption {Phase folded and binned trailed spectra.  Individual spectra were normalized to the continuum before being binned.  The grey scale is linear, with higher flux being darker.  Velocity of 0 corresponds to the rest wavelength of H$\alpha$.
}
 \label{fig:trails}
\end{figure*}

Fig.~\ref{fig:trails} displays phase-binned trailed spectra.  Each spectrum was normalized to the continuum before the data were folded and binned.  Velocities are relative to the rest wavelength of H$\alpha$.  

\section{Results}
\label{sec:results}
The spectra of all systems show the Balmer, He\,{\scriptsize I}, and  He\,{\scriptsize II} emission lines commonly detected in CVs (see Fig.~\ref{fig:lowresspec}).  In IGR J15094-6649, IGR J16500-3307, and IGR J17195-4100 the C\,{\scriptsize III}/N\,{\scriptsize III}\,$\lambda\lambda$4640--4650 Bowen blend is also seen.  

Within the uncertainty of the flux calibration of the low-resolution spectra, the 4 systems for which $UBVRI$ measurements were obtained were all at the same $V$-band brightness as when the spectra were taken a month earlier.  None of the four objects has unusual $U-B$ or $B-V$ colours (either for IPs or any other class of CVs; \citealt{BruchEngel94}).

The systems all display rapid flickering (the observational signature of mass transfer).  Three of the five targets also show coherent short-period modulations.

\subsection{Systems of unknown classification}

\subsubsection{XSS J12270-4859}
\label{sec:12270}
I did not obtain $UBVRI$ photometry or time-resolved spectroscopy of XSS J12270-4859, and the time-resolved photometry does not reveal the orbital period.
The emission lines in the low-resolution spectrum are quite weak, but, at least in the case of H$\alpha$, stronger than observed by \cite{MasettiMorelliPalazzi06}---I measure $\mathrm{EW(H\alpha)}=-21 \pm 2\,\mathrm{\AA}$, compared to the previously reported $-10.2 \pm 0.8\,\mathrm{\AA}$.

\begin{figure}
 \includegraphics[width=84mm]{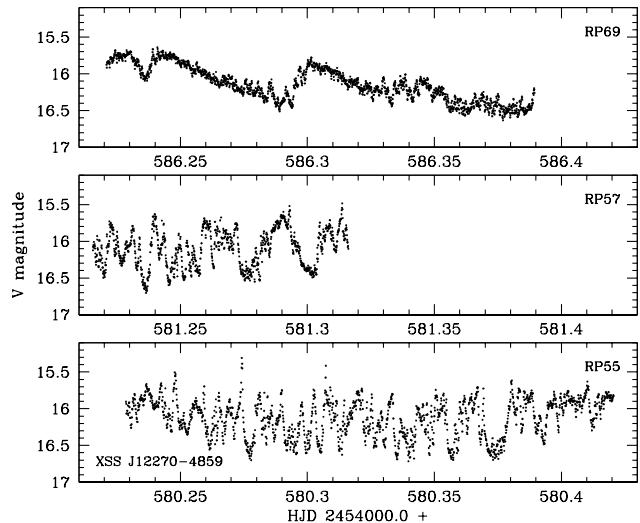}
 \caption{Light curves of XSS J12270-4859, taken with the SAAO 1-m telescope on 23, 24, and 29 April 2008.  The integration time was 10~s for run RP55, and 8~s on the other two nights.  Note the marked difference in flickering behaviour between run RP69 and the two earlier light curves.
}
 \label{fig:XSSJ12270_lc}
\end{figure}

High-speed photometry of XSS J12270-4859, covering a total of 11~h, was taken on three different nights.  The light curves, displayed in Fig.~\ref{fig:XSSJ12270_lc}, show large amplitude flickering (a range of $>1$~mag), but no periodic modulations.  Note the rather different appearance of the third run (RP69), where the largest amplitude variations are restricted to lower frequency than what is seen during the previous two observations, despite the system being at similar brightness.

\begin{figure}
 \includegraphics[width=84mm]{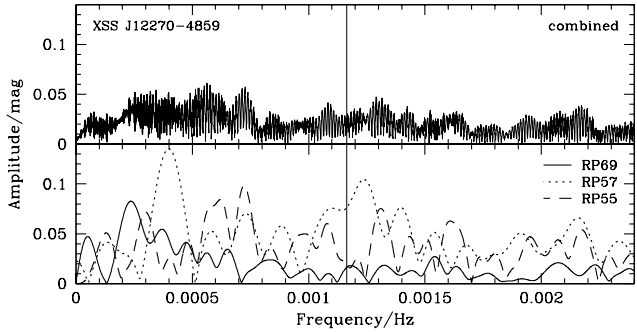}
 \caption{Fourier transforms of the light curves of XSS J12270-4859.  In the lower panel, Fourier transforms of run RP55 (dashed curve), RP57 (dotted curve), and RP59 (solid curve) are shown individually, while the Fourier transform in the upper panel is of all data combined.  The fine vertical line at 0.001163~Hz marks the frequency of the signal detected by Butters et al.\ (2008).  No significant signal appears in the optical data.
}
 \label{fig:XSSJ12270_fts}
\end{figure}

\cite{Butters08} find a signal at 859.6~s in several hours of \emph{RXTE} data of XSS J12270-4859. I can place a limit of $<0.03\,\mathrm{mag}$ on the amplitude of any optical modulation with a period at this value (see Fig.~\ref{fig:XSSJ12270_fts}; as noted above, however, the dominant optical and X-ray periods in IPs often differ).  Since quasi-coherent oscillations are common in non-magnetic CVs (see e.g.\ \citealt{Patterson81}; \citealt{WarnerWoudtPretorius03}; \citealt{Warner04}; \citealt{PretoriusWarnerWoudt06}), the period seen by \cite{Butters08} needs to be observed again before it will be clear whether it has sufficient coherence to be associated with the spin of the white dwarf.

\subsubsection{IGR J16167-4957}
\begin{figure}
 \includegraphics[width=84mm]{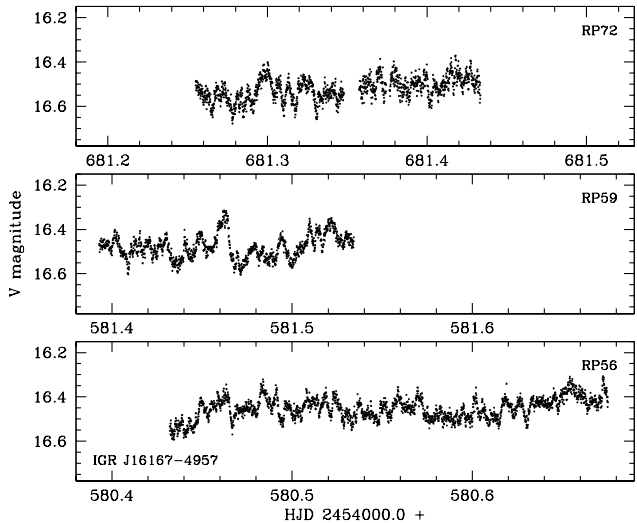}
 \caption{Light curves of IGR J16167-4957, taken at time resolution of 10~s with the SAAO 1-m and 0.76-m telescopes.  The run RP72 data were obtained under poor weather conditions.
}
 \label{fig:IGRJ16167_lc}
\end{figure}

The light curves of IGR J16167-4957 are displayed in Fig.~\ref{fig:IGRJ16167_lc}.  The 13.5~h of high-speed photometry (taken on three nights) revealed no persistent short-period modulations in the brightness of this system.  A signal at at 585~s is present in the first half of run RP56, but it is not detected in the rest of that light curve or in either of the other two light curves of this object.  This modulation was probably a quasi-periodic oscillation (e.g.\ \citealt{Warner04}).

An orbital period of $5.004 \pm 0.005\,\mathrm{h}$ is measured from the time-resolved spectroscopy of IGR J16167-4957.  This signal is not detected in the light curves.  The phase-binned trailed spectrum shows a pure S-wave modulation (see Fig.~\ref{fig:trails}). 

There is only one polar known with a period above 5~h (V1309 Ori; see \citealt{Garnavich94}); it would therefore be surprising if IGR J16167-4957 turned out to be a polar.

\subsection{Systems classified as intermediate polars}
The remaining three targets (IGR J15094-6649, IGR J16500-3307, and IGR J17195-4100) are confirmed as IPs.  

\begin{figure*}
 \includegraphics[width=168mm]{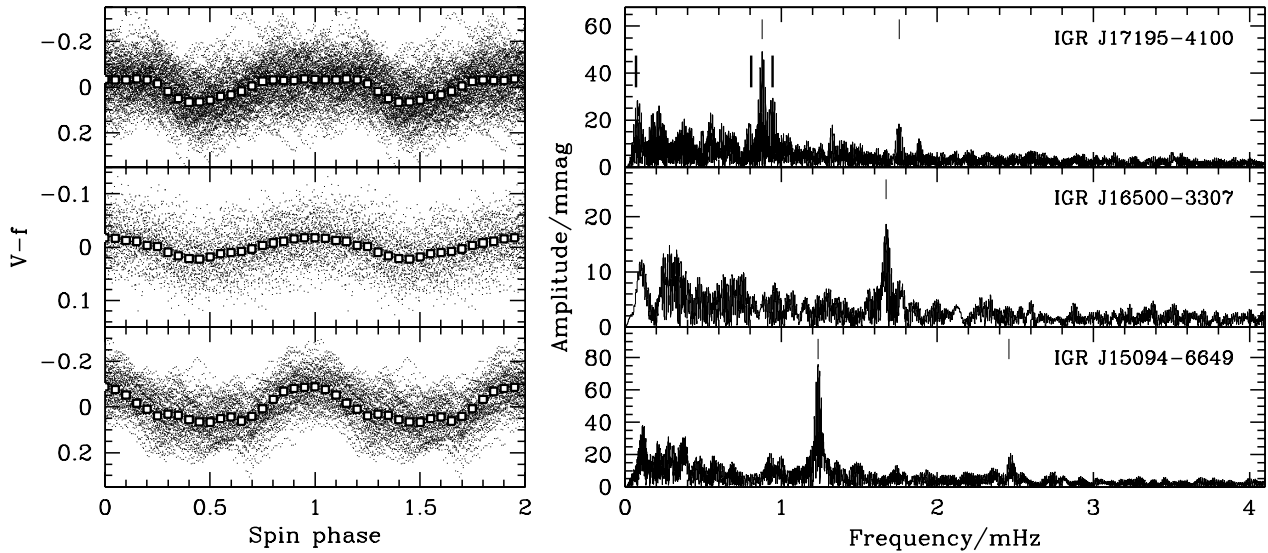}
 \caption {Left-hand panels: the spin phase folded photometry of IGR J15094-6649, IGR J16500-3307, and IGR J17195-4100.  Small points are all the data, and large open points are binned photometry.  Low-order polynomial fits were subtracted from the light curves before the data were folded and binned, and before Fourier transforms were calculated.  Right-hand panels: Fourier transforms of the high-speed photometry obtained for each system.  The orbital frequency and two orbital side bands ($\omega-\Omega$ and $\omega+\Omega$) are indicated by bold vertical bars in the Fourier transform of IGR J17195-4100 (although $\omega-\Omega$ does not correspond to a significant signal in the data; see below).  Fine vertical bars mark $\omega$ in all three Fourier transforms, and $2\omega$ in the case of IGR J15094-6649 and IGR J17195-4100.  These plots include all the data for IGR J15094-6649 and IGR J17195-4100, but only three of the runs on IGR J16500-3307.
}
 \label{fig:phifold}
\end{figure*}

Fourier transforms of the light curves of these three CVs are displayed in the right-hand panels of Fig.~\ref{fig:phifold}.  X-ray data may in future show that the strongest optical signal is an orbital side band, but I will in each case interpret it as the white dwarf spin modulation.  These frequencies (and, for IGR J15094-6649 and IGR J17195-4100, the first harmonics) are marked by fine vertical bars.  The left-hand panels of Fig.~\ref{fig:phifold} show the photometry of the new IPs, folded on the spin periods.

\subsubsection{IGR J15094-6649}
IGR J15094-6649 was significantly redder than during the discovery observations of \cite{MasettiMorelliPalazzi06} when the low-resolution spectrum was taken.  The $UBVRI$ photometry, taken 30 days later, shows a blue continuum again, although the system is at the same brightness (see Fig.~\ref{fig:lowresspec}).  The He\,{\scriptsize II}\,$\lambda$4686 emission line is particularly strong in this CV.

\begin{figure}
 \includegraphics[width=84mm]{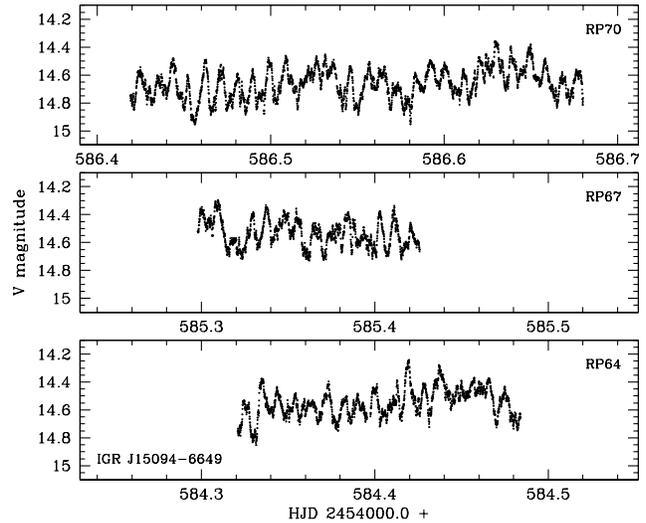}
 \caption{The three longest light curves of IGR J15094-6649.  These data were taken at a time resolution of 7~s, with the SAAO 1-m telescope, on three consecutive nights.  The white dwarf spin modulation can easily be picked out by eye.
}
 \label{fig:IGRJ15094_lc}
\end{figure}

High-speed photometric observations were obtained on four consecutive nights; three of the light curves are shown in Fig.~\ref{fig:IGRJ15094_lc}.  A prominent periodic oscillation can easily be seen in the photometry.  Assuming that this oscillation represents the white dwarf spin, IGR J15094-6649 has $P_1=13.4904 \pm 0.0003\,\mathrm{min}$.  The white dwarf spin modulation is significantly non-sinusoidal, as indicated by the detection of the first harmonic of the spin frequency (see Fig.~\ref{fig:phifold}).

Time-resolved spectroscopy was taken on three nights, and the radial velocities reveal an orbital period of $5.89 \pm 0.01\,\mathrm{h}$ (see Fig.~\ref{fig:rvandfts}).  This period does not appear in the photometry.  The trailed spectrum in Fig.~\ref{fig:trails} shows that the H$\alpha$ emission line profile varies with orbital phase.

\subsubsection{IGR J16500-3307}
\begin{figure}
 \includegraphics[width=84mm]{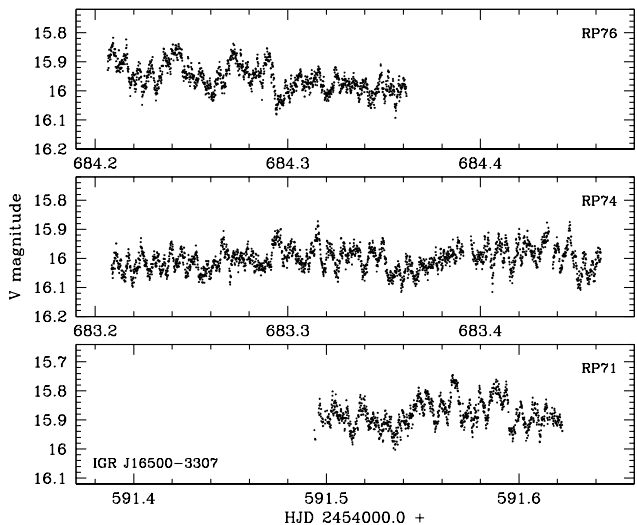}
 \caption{The three longest light curves of IGR J16500-3307, all taken at 10-s time resolution with the SAAO 0.76-m telescope.  The periodic oscillations are not very prominent, but can be seen directly in the light curves.
}
 \label{fig:IGRJ16500_lc}
\end{figure}

IGR J16500-3307 was near $V=16$ during all my observations.  The three longest light curves of this system are displayed in Fig.~\ref{fig:IGRJ16500_lc}. 

The timing of the high-speed photometric observations was not ideal---data were taken on three nights spread over a 7 day baseline, and on another 3 nights, almost 3 months later.  An oscillation with a period near 10~min is detected in all the light curves, but the cycle count between different runs is ambiguous, because of the long total baseline and poor sampling.  Using only the three light curves taken on consecutive nights in August 2008, I measure $P_1=9.9653 \pm  0.0007\,\mathrm{min}$.  The spin phase folded photometry and the Fourier transform in Fig.~\ref{fig:phifold} show data from these three nights.  No other significant periodicity is detected in the photometry of IGR J16500-3307.

I measure an orbital period of $3.617 \pm 0.003\,\mathrm{h}$ from time-resolved spectroscopy taken on 4 consecutive nights (see the phase-folded radial velocity curve and Fourier transform in Fig.\ref{fig:rvandfts}).  The trailed spectrum (Fig.~\ref{fig:trails}) shows only an S-wave modulation.

\subsubsection{IGR J17195-4100}
\label{sec:17195}
\begin{figure}
 \includegraphics[width=84mm]{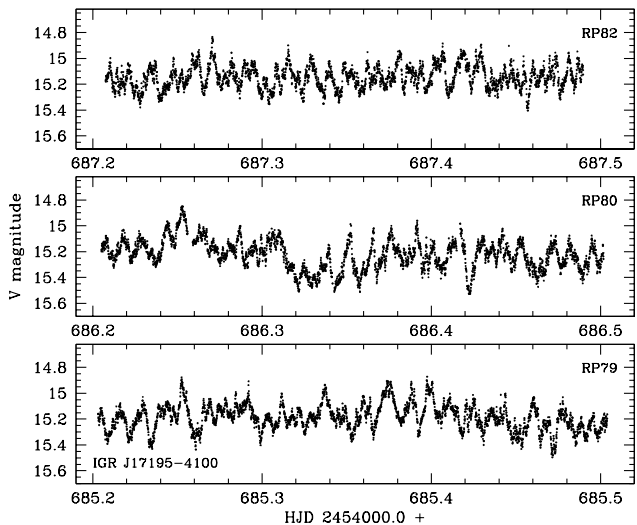}
 \caption{The three longest light curves of IGR J17195-4100.  All photometry of this system was obtained with the SAAO 0.76-m telescope, using 8-s exposures.  The modulation interpreted as the white dwarf spin cycle is the most obvious variation visible in the photometry.
}
 \label{fig:IGRJ17195_lc}
\end{figure}

I obtained a total of 32.3~h of high-speed photometry of IGR J17195-4100 on 6 consecutive nights in August 2008.  Three of the light curves are shown in Fig.~\ref{fig:IGRJ17195_lc}.  A prominent oscillation with a period near $19$~min is present in the data.  Combining all the time-resolved photometry and assuming that the largest amplitude signal is the white dwarf spin modulation, I measure $P_1=18.9925 \pm 0.0006$~min; the first harmonic of the spin frequency is also detected (see Fig.~\ref{fig:phifold}).  

Time-resolved spectroscopy of IGR J17195-4100 yields $P_{orb}=4.005 \pm 0.006\,\mathrm{h}$.  The radial velocity amplitude is low (only $28\,\mathrm{km/s}$; see Fig.~\ref{fig:rvandfts} and \ref{fig:trails}).

\begin{figure}
 \includegraphics[width=84mm]{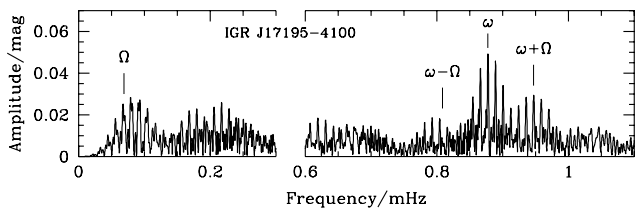}
 \caption{An expanded view of the Fourier transform of the IGR J17195-4100 photometry.  The orbital and spin frequency, and the frequencies of two orbital side bands ($\omega-\Omega$ and $\omega + \Omega$) are marked.  A spike appears near $\Omega$, and a signal at $\omega + \Omega$ is clearly detected, but, although a band of power is present near $\omega-\Omega$, this side band frequency does not correspond to any significant modulation in the data.
}
 \label{fig:IGRJ17195_ftexpand}
\end{figure}

A low-amplitude photometric signal with a period near $P_{orb}$ appears in the photometry.  A band of power is also present near $\omega-\Omega$, but this frequency cannot be identified with a spike of significant amplitude in the Fourier transform (see Fig.~\ref{fig:phifold}, as well as the more detailed Fig.~\ref{fig:IGRJ17195_ftexpand}).  The orbital side band at $\omega+\Omega$, on the other hand, is clearly detected.  Note, however, that these optical data do not reliably distinguish between the white dwarf spin frequency and orbital side bands---it is possible that the frequencies I have assumed to be $\omega$ and $\omega+\Omega$ are in fact $\omega-\Omega$ and $\omega$, respectively.

\cite{Butters08} report candidate periods of 1842~s and 2645~s in this source; these are both incommensurate with the significant frequencies I find.  

\section{Discussion}
\label{sec:disc}
\emph{INTEGRAL} has to date detected 27 CVs, of which 14 were not previously known.  This is a particularly interesting CV sample---2 of the 6 known asynchronous polars are included, and several systems were known to be IPs, or have been shown to be IPs since their detection by \emph{INTEGRAL}.  As a result of finding a large fraction of magnetic CVs in soft $\gamma$-rays, many CVs identified via \emph{INTEGRAL} detections are immediately argued to be IPs, in the absence of convincing evidence for such a classification.

For example, it has been suggested that their detection by \emph{INTEGRAL}, together with the strengths of their optical emission lines, indicate that the 5 CVs included the present study are all IPs.  However, these observational properties do not definitively point to an IP nature.  The very well-studied dwarf nova SS Cyg is an \emph{INTEGRAL} source, and so are several polars.  Also, the presence and strength of high-excitation lines are not of much use in determining the subtype of a CV.  Many non-magnetic CVs display He\,{\scriptsize II}\,$\lambda$4686 emission, and there is no clear division between magnetic and non-magnetic CVs in terms of either EW(He\,{\scriptsize II}\,$\lambda$4686) or He\,{\scriptsize II}\,$\lambda$4686/H$\beta$ line ratio.  Until the white dwarf spin modulation (or evidence of cyclotron emission) is detected, an IP classification is tentative at best.  My high-speed optical photometry confirms only 3 out of the 5 systems included in this study as IPs, bringing the total of confirmed IPs that are \emph{INTEGRAL} sources to 12.

The new IPs all have orbital periods above the period gap, where most CVs in this class are found.

It has long been known (see e.g.\ \citealt{Patterson94}) that the fraction of IPs increases in CV samples selected at shorter wavelengths (as is expected from their hard X-ray spectra and large optical to X-ray flux ratios). IPs are rare in optical and soft X-ray selected samples---in the \cite{rkcat} sample of CVs with known orbital periods (a mainly optically selected sample) roughly 7\% are IPs, and $\simeq$12\% of CVs detected at 0.1 to 2.4~keV (in the \emph{ROSAT} Bright Survey; \citealt{SchwopeBrunnerBuckley02}) are IPs.  At higher energies, however, the fraction of IPs increases.  $\simeq$40\% and  $\sim$50\% of CVs detected in the 1 to 20~keV band (by the \emph{HEAO-1} satellite; \citealt{Silber92}) and the 20 to 100~keV \emph{INTEGRAL} band, respectively, are IPs.

Any study hoping to characterize the CV population detected in soft $\gamma$-rays will require reliable classifications, as well as orbital period measurements and, in the case of IPs, white dwarf spin period measurements.  Most of the data needed can easily be obtained using small telescopes, and such follow-up may well lead to the recognition of unusual non-magnetic CVs or new asynchronous polars.

\section{Summary}
\label{sec:summ}
I have presented optical spectroscopic and photometric observations of five CVs detected by \emph{INTEGRAL}.  It has previously been proposed that all these systems systems are IPs, but I confirm only three of the classifications. 

White dwarf spin periods of 13.4904, 9.9653, and 18.9925~min are measured for IGR J15094-6649, IGR J16500-3307, and IGR J17195-4100, respectively from high-speed optical photometry.  Phase-resolved spectroscopy has revealed orbital periods for four systems; the measurements are 5.89, 5.004, 3.617, and 4.005~h for IGR J15094-6649, IGR J16167-4957, IGR J16500-3307, and IGR J17195-4100, respectively.  The orbital and spin periods are also listed together in Table~\ref{tab:targs}.

No periodic modulations were detected in the optical light curves of XSS J12270-4859 and IGR J16167-4957.  If the 859.6-s period seen in X-ray data of XSS J12270-4859 \citep{Butters08} is shown to be coherent, it will mean that this system is an IP.  Meanwhile, classification of either XSS J12270-4859 or IGR J16167-4957 as an IP is not well-grounded.

\section*{Acknowledgments}
I thank Brian Warner and Christian Knigge for helpful comments on a first draft of this paper.

\bsp

\label{lastpage}

\end{document}